\documentstyle[psfig]{mn}
\begin{document}
\newcommand{\lsim}{\mathrel{\rlap{\raise -.3ex\hbox{${\scriptstyle\sim}$}}%
                   \raise .6ex\hbox{${\scriptstyle <}$}}}%
\newcommand{\gsim}{\mathrel{\rlap{\raise -.3ex\hbox{${\scriptstyle\sim}$}}%
                   \raise .6ex\hbox{${\scriptstyle >}$}}}%
\def\simlt{\mathrel{\rlap{\lower 3pt\hbox{$\sim$}}
        \raise 2.0pt\hbox{$<$}}}
\def\simgt{\mathrel{\rlap{\lower 3pt\hbox{$\sim$}}
        \raise 2.0pt\hbox{$>$}}}

\title[Confusion noise at far-IR to millimeter
wavelengths] {Confusion noise at far-IR to millimeter wavelengths}

\author[M. Negrello, M. Magliocchetti,
        L. Moscardini, G. De Zotti, G.L. Granato, L. Silva]
{\parbox[t]{\textwidth} {M.~Negrello$^{1}$,
M.~Magliocchetti$^{1}$, L.~Moscardini$^{2}$, G.~De Zotti$^{1,3}$,
G.L.~Granato$^{1,3}$, L.~Silva$^{4}$}
\vspace*{6pt} \\
$~$ \\
$^1$SISSA, Via Beirut 4, I-34014, Trieste, Italy \\
$^2$Dipartimento di Astronomia,
Unversit\`a di Bologna, Via Ranzani 1, I-40127 Bologna, Italy \\
$^3$INAF -- Osservatorio Astronomico
di Padova, Vicolo dell'Osservatorio 5, I-35122 Padova, Italy \\
$^4$INAF -- Osservatorio Astronomico di Trieste, Via G.B. Tiepolo
11, I-34131 Trieste, Italy \\ }

\maketitle
\begin{abstract}
We present detailed predictions for the confusion noise due to
extragalactic sources in the far-IR/(sub)-millimeter channels of
ESA/ISO, NASA/Spitzer, ESA/Herschel and ESA/Planck satellites,
including the contribution from clustering of unresolved SCUBA
galaxies.  Clustering is found to increase the confusion noise,
compared to the case of purely Poisson fluctuations, by 10--15\% for
the lowest frequency (i.e. lowest angular resolution) Spitzer and
Herschel channels, by 25--35\% for the $175\,\mu$m ISOPHOT channel,
and to dominate in the case of Planck/HFI channels at $\nu \ge
143\,$GHz. Although our calculations make use of a specific
evolutionary model (Granato et al. 2004), the results are strongly
constrained by the observed counts and by data on the redshift
distribution of SCUBA sources, and therefore are not expected to be
heavily model dependent. The main uncertainty arises from the poor
observational definition of the source clustering properties. Two
models have been used for the latter: a power-law with constant slope
and a redshift-independent comoving correlation length, $r_0$, and the
standard theoretical model for clustering evolution in a $\Lambda$CDM
universe, with a redshift-dependent bias factor. In both cases, the
clustering amplitude has been normalized to yield a unit angular
correlation function at $\theta_0=1''$--$2''$ for $850\mu$m sources
fainter than 2 mJy, consistent with the results by Peacock et
al. (2000). This normalization yields, for the first model, $r_{0}\sim
8.3$ Mpc/h, and, for the second model, an effective mass of dark
matter haloes in which these sources reside of $M_{\rm
halo}\sim1.8\times10^{13}$ M$_{\odot}$/h. These results are consistent
with independent estimates for SCUBA galaxies and for other, likely
related, sources.

\end{abstract}

%
%%%%%%%%%%%%%%%%%%%%%%%%%%%%%%%%%%%%%%%%%%%%%%%%%%%%%%%%%%%%%%%%%%%%%%%%%
%
\section{Introduction}

Our knowledge of the early galaxy evolution has significantly improved
in the last years thanks to the observations performed in the
sub-millimeter waveband, in particular with the 450-850$\mu$m
Sub-millimeter Common User Bolometric Array (SCUBA) camera (Holland et
al. 1999). The sources detected by SCUBA likely represent the
progenitors of the local giant elliptical galaxies whose properties
indicate that the bulk of their star formation activity occurred at
high redshift and in a relatively short time (Thomas et al. 2002;
Romano et al. 2002). SCUBA galaxies are in fact located at high
($\gsim 2$) redshifts (Dunlop 2001; Ivison et al. 2002; Chapman et
al. 2002, 2003) and display an enormous star formation rate
($\sim$1000 M$_{\odot}$/yr), which allows masses in stars of the order
of 10$^{11}$ M$_{\odot}$ to be assembled in times shorter than 1 Gyr
(Smail et al. 2002; Aretxaga et al. 2003; Chapman et al. 2003; Saracco
et al. 2003).

For dark matter to stellar mass ratios typical of massive ellipticals
(see, e.g., Marinoni \& Hudson 2002; McKay et al.  2002), the dark
matter haloes in which the sub-millimeter galaxies reside have masses
of $\geq 10^{13}$ M$_{\odot}$. Since these massive haloes sample the
rare high-density peaks of the primordial dark matter distribution
(Kaiser 1984; Mo \& White 1996), SCUBA sources are expected to
exhibit a strong spatial clustering, similar to that measured for
Extremely Red Objects (EROs; Daddi et al. 2001, 2003), for which
similar masses have been inferred (Moustakas\& Somerville 2002).

Direct measurements of clustering properties of SCUBA galaxies are
made difficult by the poor statistics and by the fact that they
are spread over a wide redshift range, so that their clustering
signal is strongly diluted. However, tentative evidences of strong
clustering with a correlation length of $\sim 8$--13 Mpc/h,
consistent with that found for EROs, have been reported (Webb et
al. 2003; Smail et al. 2003). Peacock et al. (2000, henceforth
P00), after having removed all sources brighter than 2 mJy, found,
in the $\sim2\times2$ arcmin$^{2}$ 850$\mu$m map by Hughes et al.
(1998) in the Hubble Deep Field, some evidence for clustering of
the background source population consistent with an angular
correlation function of the form
$w(\theta)=(\theta/\theta_0)^{-0.8}$ with $\theta_0 = 1''$--$2''$.

As first pointed out by Scott \& White (1999), the clustering signal
may provide the dominant contribution to the power spectrum of
small-scale fluctuations due to extragalactic sources at
(sub)-millimeter wavelengths. Detailed calculations were carried out
by Haiman \& Knox (1999), Knox et al. (2001), Magliocchetti et al.
(2001), and Perrotta et al. (2003). The general conclusion is that the
amplitude of the clustering signal on sub-degree scales is comparable
to that from the cosmic microwave background at $850\,\mu$m, and
quickly overwhelms it at shorter wavelengths.

We address this issue from a different point of view, i.e. we analyze
the effect of clustering in increasing the confusion noise (and
therefore the detection limit) in far-IR/sub-millimeter surveys. The
effect is particularly relevant for confusion limited surveys with
poor spatial resolution, since, for small enough angular scales, the
ratio of Poisson-to-clustering fluctuations decreases with increasing
angular scale (De Zotti et al. 1996).

In particular, we consider the surveys in the ISOPHOT $175\,\mu$m
channel, in the Spitzer (formerly SIRTF) MIPS $160\,\mu$m channel,
and those performed with Herschel/SPIRE, and with Planck/HFI.
Essentially all current estimates of the confusion noise for these
instruments (Blain et al. 1998; Dole et al. 2001; Xu et al. 2001;
Franceschini et al. 2001; Rowan-Robinson 2001; Lonsdale et al.
2003; Lagache et al. 2003; Dole et al. 2003; Rodighiero et al.
2003) do in fact only take into account the Poisson contribution.
An attempt to allow also for the effect of clustering was worked
out by Toffolatti et al. (1998), with reference to the Planck
mission.

For our calculations we adopt the model by Granato et al. (2004,
henceforth GDS04), which successfully reproduces the observed
counts and the available information on the redshift distribution
of SCUBA sources. The data actually tightly constrain the volume
emissivity of these sources as a function of cosmic time, one of
the key ingredients to derive the amplitude of the fluctuations.
As mentioned above, the other key ingredient, namely the
correlation function, is only poorly known, although some
observational hints exist.

The layout of the paper is as follows. In Section~\ref{sec:formalism}
we describe the formalism.  Section~\ref{sec:model} summarizes the
main aspects of the model by Granato et al. (2004). In
Section~\ref{sec:clustering} we illustrate the models for the
two-point spatial correlation function of SCUBA sources. In Section
\ref{sec:results} we present and discuss our main results.

Throughout the paper we will assume a flat cold dark matter (CDM)
cosmology with $\Omega_{\Lambda}=0.7$ and
$\hbox{h}=H_0/100\,\hbox{km}\,\hbox{s}^{-1}\,\hbox{Mpc}^{-1}=0.7$,
in agreement with the first-year WMAP results (Spergel et al.
2003).

%
%%%%%%%%%%%%%%%%%%%%%%%%%%%%%%%%%%%%%%%%%%%%%%%%%%%%%%%%%%%%%%%%%%%%%
%

\section{Confusion noise due to clustered sources: the formalism}
\label{sec:formalism}

The autocorrelation function, $C(S_d,\theta_{\star})$, of
intensity fluctuations due to sources fainter than the detection
limit $S_d$, as a function of the angular separation
$\theta_{\star}$, writes (see De Zotti et al. 1996):
\begin{eqnarray}
C(S_d,\theta_{\star}) =
\int_{{\cal{Z}}}dz_{1}dz_{2}\int_{{\cal{L}}}dL_{1}dL_{2}
\frac{L_{1}K(z_{1})}{4\pi
d^{2}_{L}(z_{1})}\frac{L_{2}K(z_{2})}{4\pi d^{2}_{L}(z_{2})}\times \nonumber \\
\int_{\rm beam} \!\!\!\!\!\!\!\!
d\Omega_{1}d\Omega_{2}\langle\delta{\cal{N}}(L_{1},{\bf{r_{1}}})
\delta{\cal{N}}(L_{2},{\bf{r_{2}}})\rangle
f(\theta_{1},\phi_{1})f(\theta_{2},\phi_{2})\ . \label{eq:Cth}
\end{eqnarray}
Here $d_{L}(z)$ is the luminosity distance,
$K(z)=(1+z)L[\nu(1+z)]/L[\nu]$ is the K-correction for monochromatic
observations at the frequency $\nu$, the vectors ${\bf{r_1}}$ and
${\bf{r_2}}$ represent the two directions of observation separated by
an angle $\theta_{\star}$, and $\delta{\cal{N}}(L,{\bf{r}})$ is the
fluctuation around the mean number density ${\cal{N}}(L,z)$ of sources
with luminosity $L$ and redshift $z$:
\begin{eqnarray}
{\cal{N}}(L,z) \equiv \frac{dV}{dzd\Omega}(z)\Phi(L,z)\ ,
\end{eqnarray}
$dV/dzd\Omega$ being the comoving volume element, and $\Phi$ the
luminosity function. The angles ($\theta_1$, $\phi_1$) and
($\theta_2$, $\phi_2$) are measured from the axes ${\bf{r_1}}$ and
${\bf{r_2}}$, respectively.

The integration in redshift is made within the interval
$\cal{Z}=$[$z_{\rm min},z_{\rm max}$], where the sources are expected
to exist. The integration over $L$ runs over the interval
$\cal{L}=$[$L_{\rm min},\min[L_{\max},L(S_d,z)]$], where $L_{\rm max}$
and $L_{\rm min}$ are the maximum and minimum intrinsic luminosities,
and $L(S_d,z)$ corresponds to the flux density $S_d$ at the redshift
$z$.

The spatial response function of the detector, $f(\theta,\phi)$,
is assumed to be axially symmetric and Gaussian:
\begin{equation}
f(\theta,\phi)=e^{-(\theta/\Theta)^{2}/2}\ ,
\label{eq:response}
\end{equation}
where $\Theta$ relates to the beam FWHM through
\begin{eqnarray}
\Theta=\frac{\rm{FWHM}}{2\sqrt{2\ln2}}\ .
\end{eqnarray}
If the luminosities of the sources are
statistically independent of their positions, we get (Dautcourt 1977):
\begin{eqnarray}
<\delta{\cal{N}}(L_{1},{\bf{r_{1}}})
\delta{\cal{N}}(L_{2},{\bf{r_{2}}})>~=~~~~~~~~~~~~~~~~~~~~~~~~~~~~~ \nonumber \\
~~~~~~~~~~~~~~~~~~~~~~~~~ {\cal{N}}(L_{1},z_{1})\delta^{D}({\bf{r_{2}-r_{1}}})\delta^{D}(L_{2}-L_{1})
\nonumber \\
~~~~~~~~~~~~~~~~~~~~~~~~~+~{\cal{N}}(L_{1},z_{1}){\cal{N}}(L_{1},z_{2})\xi({\bf{r_{2}-r_{1}}},z)\ ,
\label{eq:clust}
\end{eqnarray}
$\delta^{D}$ being the Dirac function. The first term on the
right-hand side corresponds to a Poisson distribution of point
sources, while the second one is due to clustering with a two-point
spatial correlation function $\xi$.

The variance of intensity fluctuations within the telescope beam,
whose square root, $\sigma_{N}$, will be referred to as {\it{confusion
noise}}, is obtained setting $\theta_{\star}=0$:
\begin{eqnarray}
\sigma^{2}_{N}(S_d) = \lim_{\theta_{\star}\rightarrow0}
C(S_d,\theta_{\star}) = \sigma^{2}_{P}+\sigma^{2}_{C}\ ,
\label{eq:conf_noise}
\end{eqnarray}
and is the quadratic
sum of the Poisson term:
\begin{eqnarray}
\sigma^{2}_{P} &=& \int_{beam}d\Omega f^{2}(\theta,\phi)\times  \nonumber \\
&~& \int_{\cal{Z}}dz\frac{dV}{dzd\Omega}(z)
\int_{\cal{L}}dL\left(\frac{LK(z)}{4\pi
d^{2}_{L}(z)}\right)^{2}\Phi(L,z)\ ,
\label{eq:sigma_poiss}
\end{eqnarray}
and of the clustering contribution, which, in the small angle
approximation (i.e. $\theta_1,\theta_2 << 1$) and assuming that the
maximum scale of appreciable clustering is much smaller than the
Hubble radius, $c/H_0$, writes:
\begin{eqnarray}
\sigma^{2}_{C}=\left(\frac{1}{4\pi}\frac{c}{H_0}\right)^2
\int_{beam}d\Omega_{1}d\Omega_{2}
f(\theta_1,\phi_1)f(\theta_2,\phi_2)\times ~~~~~ \nonumber \\
\int_{\cal{Z}}dz\frac{j^{2}_{\rm eff}(z,S_d)}{(1+z)^{4}E^{2}(z)}
\int_{d_{\vartheta}(z)}^{r_{\rm sup}}dr\frac{2}{c/H(z)}
\frac{\xi(r,z)}{\sqrt{1-(d_{\vartheta}(z)/r)^{2}}}\ ,
\label{eq:sigma_clust}
\end{eqnarray}
where $r$ is a comoving scale and $j_{\rm eff}(z,S_d)$ is the
effective comoving volume emissivity at redshift $z$ contributed by
sources with fluxes $S<S_{d}$:
\begin{eqnarray}
j_{\rm eff}(z,S_{d})=\int_{\cal{L}}dL\, K(z)L\Phi(L,z)\ .
\label{eq:j_eff}
\end{eqnarray}
In Eq.~(\ref{eq:sigma_clust}) $E(z)$ describes the time evolution of
the Hubble parameter [$H(z)=H_0E(z)$], $d_{\vartheta}(z)$ is the
comoving linear distance corresponding, at a given redshift $z$, to an
angular separation $\vartheta$ in the sky\footnote{In the small angle
approximation:
$\vartheta^2=\theta^2_1+\theta^2_2-2\theta_1\theta_2\cos(\phi_1-\phi_2)$}. The
upper limit of integration over $r$, $r_{\rm sup}$, is the comoving
scale at which the correlation vanishes; we adopt $r_{\rm
sup}=30\,$Mpc/h, i.e. approximately three times the measured
clustering radius $r_0$ (see Section~\ref{sec:clustering}).

The detection limit, $S_d$, for a confusion limited survey, is defined
by:
\begin{eqnarray}
S_{d}=q \times \sigma_{N}(S_{d})\ , \label{eq:Slim}
\end{eqnarray}
where the parameter $q$ is usually chosen in the range [3,5].

%
%%%%%%%%%%%%%%%%%%%%%%%%%%%%%%%%%%%%%%%%%%%%%%%%%%%%%%%%%%%%%%%%%%%
%

\section{The evolutionary model}
\label{sec:model}

According to the GDS04 model, SCUBA sources are interpreted as
spheroids observed during their major episode of star-formation, whose
evolution and duration, shorter for more massive objects, is
substantially affected not only by supernova feedback, but also by the
growth of a central super-massive black hole (SMBH), and by the
ensuing QSO activity. The relative importance of the QSO feedback,
compared to supernovae, increases with the galaxy mass.  This
scenario, already explored in a phenomenological way by Granato et
al. (2001), has been substantiated by GDS04 in a model which follows,
using simple semi-analytic prescriptions, the evolution of baryons
within dark matter haloes with total mass $M_{\rm vir} {\gsim} 2.5
\times 10^{11} M_\odot$, forming at $z
\gsim 1.5$ at the rate predicted by the canonical hierarchical
clustering scenario. The mass and redshift cuts are meant to crudely
filter out the haloes associated with disk and irregular galaxies,
whose formation is not quantitatively addressed by GDS04. However,
disk and irregular galaxies are envisaged as associated to haloes
virializing at lower redshifts, eventually incorporating most of the
less massive haloes which virialized at earlier times and may become
galactic bulges.

We note that according to GDS04 (see also Granato et al. 2001), the
high-redshift QSO activity marks and concurs to the end of the major
episode of star formation in spheroids. Thus, there is a clear
evolutionary link between SCUBA sources and high-redshift QSOs.  Also,
in the mass and redshift range considered, the model implies a
one-to-one correspondence between haloes and spheroidal galaxies,
consistent with the available data on the clustering of Lyman-break
galaxies (Bullock et al. 2002) and SCUBA galaxies (Magliocchetti et
al. 2001)

The GDS04 model predicts a local SMBH mass function and a
relationship between black hole and spheroid mass (in stars) which
is in excellent agreement with the observational data. Also, when
coupled with the most updated version of GRASIL (whose original
version is described by Silva et al. 1998), the code which
computes in a self-consistent way the chemical and
spectro-photometric evolution of galaxies from far-UV to radio, it
yields predictions which are fully consistent with a number of
observables extremely challenging for all other semi-analytic
models, including the sub-millimeter number counts and redshift
distribution of spheroidal galaxies, as well as the
epoch-dependent K-band luminosity function.

High-redshift star-forming spheroids are not the only important
class of objects in the far-infrared and sub-millimeter wavebands.
We also need to take into account late-type, normal galaxies and
starburst galaxies which in fact dominate the bright tail of the
counts. To deal with these sources, we adopt, as usual, a
phenomenological approach (e.g. Franceschini et al. 2001; Takeuchi
et al. 2001; Rowan-Robinson 2001; Gruppioni et al. 2002; Lagache
et al. 2003), which consists in simple analytic recipes to evolve
their $60\mu$m local luminosity functions (Saunders et al. 1990)
and appropriate templates for the spectral energy distributions
(SED) to extrapolate the models to different wavelengths. The
prescriptions adopted here are those defined by Silva et al.
(2004), which provide reasonable fits to the results of IRAS and
ISO surveys. Briefly, normal late-type galaxies are assumed to
undergo a pure luminosity evolution described by
$L(z)=L(0)(1+z)^{1.5}$ and to have the SED of the Sc galaxy
NGC$\,6946$. The SED of the galaxy NGC$\,6090$ is adopted for
starburst galaxies and the luminosity function of these objects,
$\Phi_{\rm SB}$, is assumed to evolve both in luminosity and in
density: $\Phi_{\rm SB}[L(z),z]=\Phi_{\rm
SB}[L(z)/(1+z)^{2.5},z=0](1+z)^{3.5}$. The evolutionary laws for
both populations were assumed to apply up to $z=1$; then the
luminosity functions were kept constant up to $z_{\rm
cutoff}=1.5$.

Spiral and starburst galaxies are relatively weakly clustered
(Fisher et al. 1994; Loveday et al. 1995; Guzzo et al. 1997;
Madgwick et al. 2003), so that their contributions to the
confusion noise turn out to be dominated by Poisson fluctuations.
In the following we will therefore neglect their contribution to
$\sigma_C$.

At millimeter wavelengths the composition of the bright counts
drastically changes since flat spectrum radio sources take over.
Their Poisson fluctuations have been estimated by extrapolating the 15
GHz differential counts, $n(S_{\rm Jy})=51S_{\rm
Jy}^{-2.15}\,\hbox{Jy}^{-1}\hbox{sr}^{-1}$ (Waldram et al. 2003), with
a spectral index equal to zero (i.e. assuming a flux density constant
with frequency), consistent with the WMAP results (Bennett et
al. 2003). This choice for the spectral index is likely to
increasingly overestimate the counts with increasing frequency, as the
radio-source spectra steepen both because emitting regions become
optically thin and because of electron ageing effects. However, this
has little impact on our results since at sub-millimeter wavelengths
radio sources are anyway a minor component.

Although radio sources are rather strongly clustered (Magliocchetti et
al. 1998; Blake \& Wall 2002a,b; Overzier et al. 2003, Magliocchetti
et al. 2004), their clustering signal is highly diluted due to the
broadness of their luminosity function (e.g. Dunlop
\& Peacock 1990) to the effect that their $\sigma_C$ is small
when compared to $\sigma_P$, and can be neglected (Arg{\" u}eso et
al. 2003; Toffolatti et al. 1998).

%
%%%%%%%%%%%%%%%%%%%%%%%%%%%%%%%%%%%%%%%%%%%%%%%%%%%%%%%%%%%%%%%%%%%%%%%
%
\section{The two-point correlation function of SCUBA galaxies}
\label{sec:clustering}

P00 carried out a power-spectrum analysis of the $850\mu$m map of the
northern {\it Hubble Deep Field} by Hughes et al.  (1998), after
subtracting sources brighter than $S_d=2\,$mJy. They found, for
multipole numbers $\ell$ in the range $10^{4}\lsim \ell
\lsim 5\times 10^{4}$, some evidence of power in excess of the sum
of instrumental noise and of estimated Poisson fluctuations due to
unresolved sources. Such excess power can be accounted for by
source clustering described by an angular correlation function of
the form:
\begin{equation}
w(\theta)=(\theta/\theta_{0})^{-0.8}\,  \label{eq:wth_2mJy_850} \ ;
\end{equation}
with $\theta_{0}$ in the range $1$--$2$ arcsec. As made clear by
P00, however, the estimated amplitude of Poisson fluctuations is
rather uncertain due to our poor knowledge of the $850\mu$m counts
below 2 mJy, and the possibility that they account entirely for
the detected confusion noise cannot be ruled out. The more recent
data on the faint $850\,\mu$m counts (Chapman et al. 2002; Smail
et al. 2002; Knudsen \& van der Werf 2003) are in good agreement
with the analysis by P00 and do not allow firmer conclusions on
the amplitude of the clustering signal. On the other hand, as
shown below, an angular correlation function with the amplitude
suggested by P00 is consistent with a number of other data on
clustering of SCUBA galaxies themselves (although, again, the
significance of the clustering detection is not very high) and of
other populations which are closely linked to such galaxies, such
as EROs. We thus conclude that Eq.~(\ref{eq:wth_2mJy_850}) with
$\theta_{0} = 1"$--$2"$ provides the current best guesstimate for
the normalization of our models.

%Even if, within the framework of the model of GDS04 that we have
%adopted to perform our calculations, we find a value $\sim$74000
%deg$^{-2}$ for the effective number counts below 2 mJy (see Eq. (20)
%of P00), not too far from the one estimated by P00 (N$_{eff}\sim92600$
%deg$^{-2}$). Regardless of the uncertainties in the clustering
%analysis presented by P00, an important point to stress is the fact
%that the angular correlation function of the background source
%population as predicted within the CDM framework is compatible (at the
%scales investigated by P00) with $\theta_{0}=1^{\prime\prime}$ if one
%assume a resonable mass of $\sim10^{13}$ M$_{\odot}$ for the dark
%matter halos hosting a typical scuba galaxy (see subsection
%\ref{subsec:model2}). This plays in favour of a non-null detection of
%clustering in the residual 850$\mu$m-map analyzed by P00 and suggests
%values for the angular correlation scale
%$\theta_{0}\simgt1^{\prime\prime}$.

The spatial correlation function, $\xi_{\rm sph}$,  can be
obtained by inverting the Limber's (1953) equation:
\begin{eqnarray}
w(\theta,S_{d}) =
\frac{1}{I_{850}^{2}(S_{d})}\int_{z_{\min}}^{z_{\max}}dz
\left(\frac{1}{4\pi}\frac{c}{H_{0}}\right)^{2}\frac{j_{\rm
eff}^{2}(z,S_{d})}{(1+z)^{4}E^{2}(z)}\nonumber \\
\times\int_{d_{\theta}(z)}^{r_{\rm sup}}dr\frac{2}{(c/H(z))}
\frac{\xi_{\rm sph}(r,z)}{\sqrt{1-(d_{\theta}(z)/r)^{2}}}\ ,
\label{eq:wth}
\end{eqnarray}
with $S_{d}=2\,$mJy. The quantity $I_{850}$ is the 850$\mu$m
background intensity produced by sources fainter than $S_d$. The other
symbols have the same meaning as in Eq.~(\ref{eq:sigma_clust}). As
mentioned in the Introduction, SCUBA galaxies are interpreted by GDS04
as forming massive spheroidal galaxies. The model yields
$I_{850}(S_d)=6.6\times10^{-19}$ erg/s/cm$^2$/Hz/sr.

\begin{figure}
\vspace{8cm} \includegraphics{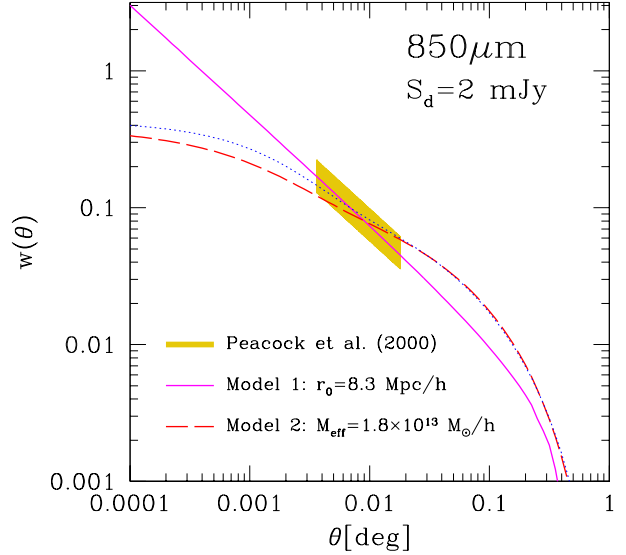} \caption{Comparison of the adopted
models for the angular correlation function with
Eq.~(\protect\ref{eq:wth_2mJy_850}) with $\theta_{0} = 1"$--$2"$, in
the interval of angular scales probed by P00. The long-dashed curve
has been obtained using the prescriptions by Peacock $\&$ Dodds (1996)
for generating the dark matter power spectrum, while the dotted curve
corresponds to the more accurate method by Smith et
al. (2003)}. \label{fig:wth_2mJy_850}
\end{figure}

In the following, we will consider both a phenomenological, power-law
model for $\xi_{\rm sph}$ (Model 1) and a physically motivated model
(Model 2). Both models are normalized to Eq.~(\ref{eq:wth_2mJy_850})
with $\theta_{0} = 1"$--$2"$ in the range of scales probed by P00 (see
Fig.~\ref{fig:wth_2mJy_850}).

\subsection{Model 1}
\label{subsec:model1}

Our first model assumes the usual power-law shape:
\begin{equation}
\xi_{\rm sph}(r,z)=[r/r_{0}(z)]^{-1.8}\ . \label{xi_pl}
\end{equation}
In view of the tight connection between spheroidal galaxies and
active nuclei at their centers, entailed by the GDS04 model, we
further assume that the correlation length, $r_0$, is constant
{\it in comoving coordinates}, as suggested by  quasar data (Croom
et al. 2001; Outram et al. 2003) at least in the redshift interval
$0.5 \leq z \leq 2.5$--3.

Then, from Eq.~(\ref{eq:wth}), we find that the range of values of
$r_0$ corresponding to $\theta_{0} = 1''$--$2''$ is:
\begin{eqnarray}
r_{0}=8.3\pm1.3 ~ {\rm Mpc}/{\rm h}\ ,
\end{eqnarray}
very close to the range of values found for bright 2QZ quasars
($r_{0}=8.37\pm1.17\,{\rm Mpc}/{\rm h}$; Croom et al. 2002), and
fully consistent with the tentative estimate of the correlation length
of SCUBA galaxies obtained by Smail et al. (2003):
$r_{0}=8\pm3\,$Mpc/{h}. The above value of the comoving correlation
length is also compatible with the ones measured for bright local
elliptical galaxies (Loveday et al.  1995; Guzzo et al. 1997; Norberg
et al. 2002; Madgwick et al. 2003), supporting the link between these
objects and the population of SCUBA sources.

\begin{table*}
\protect\small \baselineskip=12pt \noindent \centering
\caption{Estimated confusion noise and detection limits, $S_{d}$
(neglecting instrumental noise as well as emissions within our Galaxy
and fluctuations of the cosmic microwave background) for several
far-IR/(sub)-millimeter instruments. Shown are the Poisson contributions
($\sigma_P$) due to spiral galaxies (sp), star-forming galaxies (sb),
radio-galaxies (rg) and star-forming spheroids (sph). For the latter,
we also give the clustering fluctuations ($\sigma_{C,sph}$). At each
wavelength, three cases are shown: Poisson contributions only (first
line), Poisson plus clustering contribution from star-forming
spheroids with a correlation function of the form $\xi_{\rm
sph}(r,z)=(r/r_{0})^{-1.8}$ with $r_{0}=8.3\,$Mpc$/$h (Model 1,
second line), Poisson plus clustering contribution from star-forming
spheroids with a correlation function of the form $\xi_{\rm
sph}(r,z)=b^{2}(M_{\rm eff},z)\xi_{\rm DM}$ with $M_{\rm
eff}=1.8\times10^{13}~M_{\odot}/$h (Model 2, third line). We have
assumed $q=5$, so that $S_{d} = 5\times(\sigma^{2}_{P,sp}+
\sigma^{2}_{P,sb}+ \sigma^{2}_{P,rg}+ \sigma^{2}_{P,sph}+
\sigma^{2}_{C,sph})^{1/2}$.}
\begin{tabular}{ccccccccc}
\hline \hline $\lambda$ & $\nu$ & FWHM & $\sigma_{P,sp}$ &
$\sigma_{P,sb}$ & $\sigma_{P,rg}$ & $\sigma_{P,sph}$ &
$\sigma_{C,sph}$ & $S_{d}$ \\ ($\mu$m) & (GHz)
& (arcsec) & (mJy) & (mJy) & (mJy) & (mJy) & (mJy) & (mJy) \\
\hline
\multicolumn{9}{c}{{\bf{ISOPHOT}}} \\
\hline
{\bf{175}} & 1714 & 94.0 & 20 & 29 & 2 & 40 & - & 264 \\
    &      &      & 21 & 29 & 2 & 40 & 45 & 350 \\
    &      &      & 21 & 29 & 2 & 40 & 38 & 327 \\
\hline
\multicolumn{9}{c}{{\bf{Spitzer/MIPS}}} \\
\hline
{\bf{160}} & 1875 & 35.2 & 5.0 & 9.1 & 0.4 & 11.2 & - & 76.4 \\
    &      &      & 5.3 & 9.3 & 0.4 & 11.5 & 8.2 & 88.3 \\
    &      &      & 5.1 & 9.3 & 0.4 & 11.4 & 6.0 & 83.2 \\
\hline
\multicolumn{9}{c}{{\bf{Herschel/SPIRE}}} \\
\hline
{\bf{250}} & 1200 & 17.4 & 2.0 & 3.9 & 0.1 & 4.6 & - & 31.7 \\
    &      &      & 2.1 & 4.1 & 0.1 & 4.8 & 2.6 & 35.7 \\
    &      &      & 2.1 & 4.0 & 0.1 & 4.7 & 1.8 & 33.7 \\
 & & & \multicolumn{6}{c}{---------------------------------------------------------------------------------------------} \\
{\bf{350}} & 857  & 24.4 & 2.5 & 5.0 & 0.2 & 6.9 & - & 44.2 \\
    &      &      & 2.6 & 5.1 & 0.2 & 7.3 & 4.1 & 50.6 \\
    &      &      & 2.5 & 5.0 & 0.2 & 7.1 & 3.1 & 47.9 \\
 & & & \multicolumn{6}{c}{---------------------------------------------------------------------------------------------} \\
{\bf{500}} & 600  & 34.6 & 1.7 & 3.3 & 0.2 & 5.8 & - & 34.6 \\
    &      &      & 1.7 & 3.3 & 0.3 & 6.0 & 3.8 & 40.1 \\
    &      &      & 1.7 & 3.3 & 0.3 & 6.0 & 3.2 & 38.6 \\
\hline
\multicolumn{9}{c}{{\bf{Planck/HFI}}} \\
\hline
{\bf{350}} & 857 & 300 & 45 & 65 & 7 & 116 & - & 705 \\
    &     &     & 50 & 66 & 10 & 116 & 250 & 1439 \\
    &     &     & 51 & 66 & 10 & 116 & 266 & 1510 \\
 & & & \multicolumn{6}{c}{---------------------------------------------------------------------------------------------} \\
{\bf{550}} & 545 & 300 & 18 & 29 & 5 & 55 & - & 323 \\
    &     &     & 19 & 29 & 7 & 55 & 117 & 671 \\
    &     &     & 19 & 29 & 8 & 55 & 137 & 757 \\
 & & & \multicolumn{6}{c}{---------------------------------------------------------------------------------------------} \\
{\bf{850}} & 353 & 300 & 5 & 9 & 3 & 20 & - & 115 \\
    &     &     & 6 & 9 & 5 & 20 & 42 & 241 \\
    &     &     & 6 & 9 & 5 & 20 & 53 & 288 \\
 & & & \multicolumn{6}{c}{---------------------------------------------------------------------------------------------} \\
{\bf{1380}} & 217 & 300 & 1.2 & 2.2 & 1.9 & 4.9 & - & 29.2 \\
    &      &     & 1.3 & 2.2 & 2.5 & 4.9 & 10.1 & 59.0 \\
    &      &     & 1.3 & 2.2 & 2.8 & 4.9 & 13.3 & 73.2 \\
 & & & \multicolumn{6}{c}{---------------------------------------------------------------------------------------------} \\
{\bf{2100}} & 143 & 426 & 0.5 & 0.7 & 2.0 & 1.8 & - & 13.9 \\
    &      &     & 0.5 & 0.7 & 2.6 & 1.8 & 4.2 & 26.5 \\
    &      &     & 0.5 & 0.7 & 2.8 & 1.8 & 5.7 & 33.5 \\
 & & & \multicolumn{6}{c}{---------------------------------------------------------------------------------------------} \\
{\bf{3000}} & 100 & 552 & 0.2 & 0.3 & 2.5 & 0.7 & - & 12.9 \\
    &      &     & 0.3 & 0.3 & 2.8 & 0.7 & 1.8 & 17.1 \\
    &      &     & 0.3 & 0.3 & 3.0 & 0.7 & 2.6 & 20.0 \\
\hline \hline
\end{tabular}
\label{tab:results}
\end{table*}

\begin{figure*}
\centering \vspace{14.0cm} \includegraphics{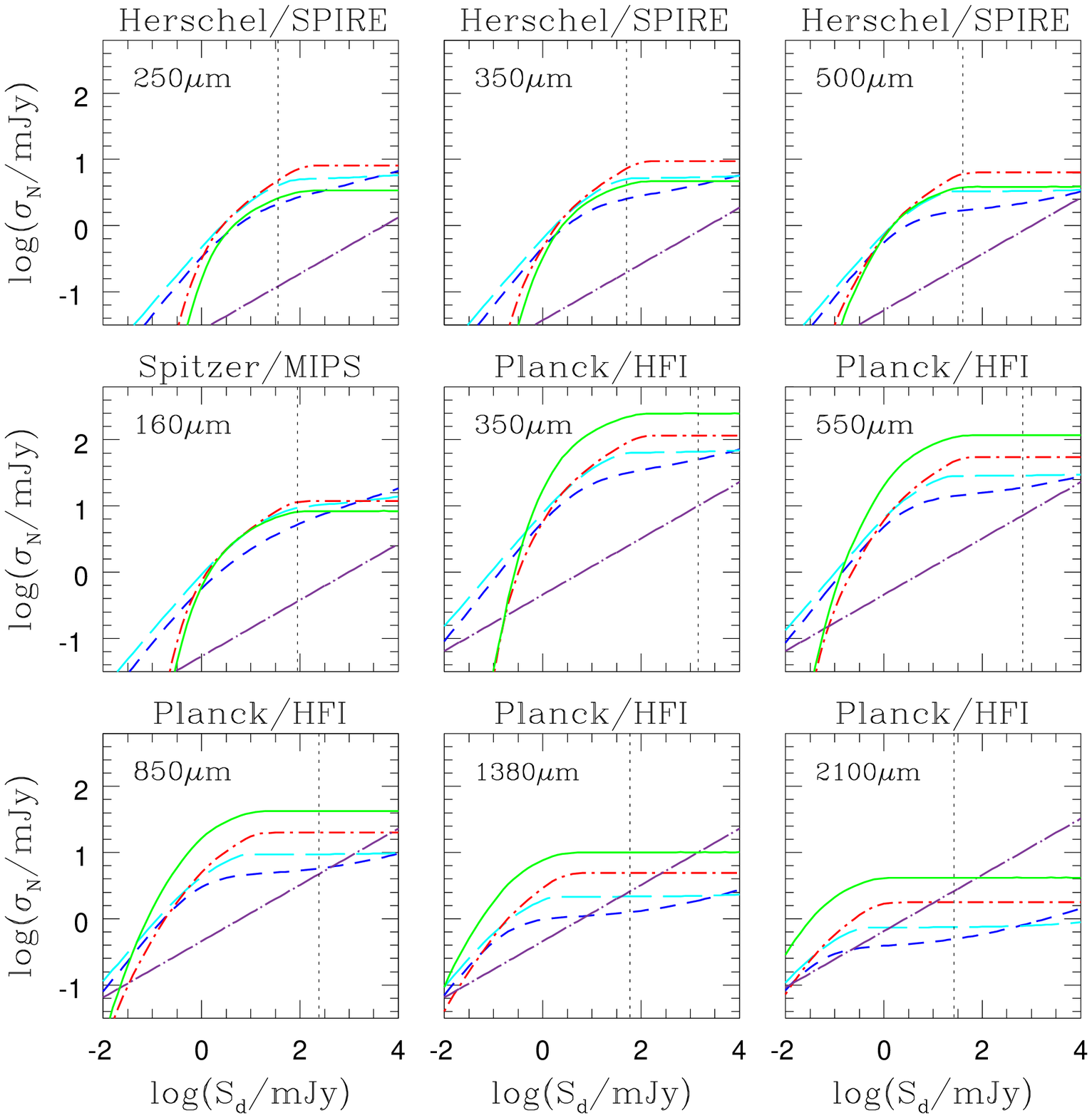} \caption{Confusion noise
$\sigma_{N}$ as a function of the detection limit $S_{d}$ for spiral
galaxies (short-dashed line), starburst galaxies (long-dashed line),
radio-sources (dotted--long-dashed line) and star-forming spheroids
(dotted--short-dashed line), all assumed to be randomly
distributed. The confusion noise due to clustered star-forming
spheroids (Model 1 with $r_{0}=8.3$ Mpc/{h}) is shown by the solid
line. The vertical dotted lines correspond to the detection limit
$S_{d}$ obtained by taking into account both Poisson and clustering
fluctuations.} \label{fig:sigma_N}
\end{figure*}

\subsection{Model 2}
\label{subsec:model2}

The spatial correlation function of visible galaxies can be described
as the product of the correlation function of dark matter, $\xi_{\rm
DM}(r,z)$, with the bias parameter which relates the distribution of
light to that of the matter (Matarrese et al.  1997; Moscardini et
al. 1998). $\xi_{\rm DM}(r,z)$ is determined by the cosmology and by
the primordial density perturbations (Peacock \& Dodds 1996; Smith et
al. 2003), for which we have adopted a CDM spectrum, with an index
$n=1$, a shape parameter $\Gamma=0.2$ and a normalization
$\sigma_8=0.8$ (see e.g. Lahav et al. 2002, Spergel et al. 2003). We
will use the practical fitting formula by Peacock $\&$ Dodds (1996) to
evolve the dark matter power spectrum into the non-linear regime. For
the cosmological model adopted here, the angular correlation function
obtained using the improved treatment by Smith et al (2003) is only
slightly different and is intermediate between those yielded by our
two models (Fig.~\ref{fig:wth_2mJy_850}).

Under the assumption of a single object per halo, which is expected to
hold for massive objects at high redshift ($z \gsim 1$) and is
implicit in the GDS04 model, the bias parameter, $b$, can be written
as a function of an effective dark-matter halo mass, $M_{\rm eff}$
(which is close to the minimum halo mass, cf.  Moscardini et
al. 1998), and of redshift (see, e.g., Mo \& White 1996; Sheth \&
Tormen 1999). We can then write the spatial correlation function of
star-forming spheroids as
\begin{equation}
\xi_{\rm sph}(r,z)=b^{2}(M_{\rm eff},z)\xi_{\rm DM}(r,z) \ .
\label{eq:xi2_obj}
\end{equation}
The effective mass $M_{\rm eff}$ is determined by the condition that,
in the range of angular scales probed by P00, the model correlation
function matches the measured one. Using the analytic formulae by
Sheth \& Tormen (1999) for the bias factor and by Peacock \& Dodds
(1996) for $\xi_{\rm DM}(r,z)$ we find:
\begin{equation}
 M_{\rm eff}=1.8\times10^{13} M_{\odot}/{\rm h} \ ,
\label{eq:Meff}
\end{equation}
consistent with the estimates by Moustakas \& Somerville (2002)
for ERO halo masses (the minimum mass of ERO dark matter haloes is
estimated to be $\sim 10^{13} M_{\odot}/{\rm h}$, while the galaxy
number-weighted average halo mass is $\simeq 5\times10^{13}
M_{\odot}/{\rm h}$). We recall that EROs are either massive dusty
galaxies like SCUBA galaxies, or evolved giant ellipticals, a
later evolutionary phase of SCUBA galaxies, in the scenario
discussed by GDS04.

For a virialization redshift $z_{\rm vir} \lsim 4$, GDS04 predict a
ratio $M_{\rm eff}/M_{\rm sph}\sim40$, where $M_{\rm sph}$ is the
present-day mass in stars of spheroids (see Figure 5 of their
paper). Thus $M_{\rm eff}$ corresponds to a mass in stars of $\simeq
6\times10^{11}$ M$_{\odot}$, compatible with values for the stellar
masses inferred for SCUBA galaxies (Smail et al. 2003) and for
high-$z$ galaxies with very red near-IR colours (Saracco et al.
2003).

$~$

As a last remark, we note that (as also illustrated by
Fig.~\ref{fig:wth_2mJy_850}) Models 1 and 2 widely differ on small
angular scales. The small scale flattening of $w(\theta)$ implied by
Model 2 is due to the fact that at high redshift ($z{\gsim}2$) density
perturbations are close to the linear regime, even on very small scales.

%
%%%%%%%%%%%%%%%%%%%%%%%%%%%%%%%%%%%%%%%%%%%%%%%%%%%%%%%%%%%%%%%%%
%

\section{Results and discussion}
\label{sec:results}

Table~\ref{tab:results} gives our estimated contributions to the
confusion noise and the corrisponding detection limits $S_d$ for all
the populations previously described (obtained by adopting $q=5$ in
Eq.~(\ref{eq:Slim})), in the case of:
\begin{itemize}
\item [i)] the $175\,\mu$m channel of the Imaging
Photo-Polarimeter of the ESA's ISO satellite (ISOPHOT);
\item [ii)] the longest wavelength channel ($160\,\mu$m) of the
Multiband Imaging Photometer (MIPS) of NASA's Spitzer (formerly
SIRTF) satellite, launched last August;
\item [iii)] all channels (250, 350 and 500$\,\mu$m) of
the Spectral and Photometric Imaging REceiver (SPIRE) of the ESA's
Herschel satellite scheduled for launch in 2007;
\item [iv)] all channels
(350, 550, 850, 1380, 2100, and $3000\,\mu$m) of the High Frequency
Instrument (HFI) of the ESA's Planck satellite, to be launched jointly
with Herschel.
\end{itemize}
The values for the angular resolutions (FWHM) have been taken from
Dole et al. (2001; ISOPHOT), Lonsdale et al. (2003; MIPS/Spitzer),
Griffin et al. (2000; SPIRE/Herschel), and Lamarre et al. (2003;
HFI/Planck).

As discussed above, forming spheroidal galaxies are the only
population, among those considered here, whose fluctuations are
dominated by clustering. We have therefore only presented in
Table~\ref{tab:results} results for the instruments and the
wavelengths where this population gives an important contribution
to the confusion noise. For each channel, the first line
corresponds to pure Poisson fluctuations while the second and the
third lines include the contributions of clustering based on Model
1 and 2, respectively. The increment of Poisson fluctuations when
we allow for the effect of clustering is obviously due to the
increment of $S_d$.

The relative importance of contributions from each class of
objects as a function of the flux limit is illustrated in
Fig.~\ref{fig:sigma_N} by the function $\sigma_N$ [see
Eq.~(\ref{eq:conf_noise})]. The flattening of the function
$\sigma_{C,sph}(S_d)$ at bright flux densities follows from the
fact that the main contribution does not come from sources just
below the detection limit, as is frequently the case for Poisson
fluctuations, but from sources at redshifts where the effective
volume emissivity, $j_{\rm eff}$ [see Eq.~(\ref{eq:sigma_clust})],
is maximum.

For power-law differential counts, $n(S)\propto S^{-\beta}$,
$\sigma_P\propto S_d^{(3-\beta)/2}$ if $\beta < 3$. This power-law
behaviour, obeyed by $\sigma_{P,rg}$ because of our adoption of the
power-law representation of the counts by Waldram et al.  (2003), must
break down at faint flux densities, where counts must converge, and
must flatten at bright flux densities ($S\gsim 1\,$Jy), where the
slope of the counts approaches the Euclidean value (Bennett et
al. 2003). Since the survey by Waldram et al.  (2003) covers the flux
density range 10--1000 mJy, we may have somewhat overestimated
fluctuations due to radio sources, if the counts start converging not
far below 10 mJy.

The slope of the bright counts of normal late-type galaxies is
close to the Euclidean value ($\beta \simeq 2.5$), while that of
starburst galaxies is somewhat steeper, due to their relatively
strong evolution. Correspondingly, for both populations $\sigma_N$
is a flatter function of $S_d$ than in the case of radio galaxies,
and is flatter for starburst than for normal galaxies.

On the other hand, the bright tail of the counts of forming
spheroidal galaxies is extremely steep ($\beta > 3$), as a
consequence of the combined effect of strong evolution and
negative K-correction. Thus, the main contribution to
$\sigma_{P,sph}(S_d)$ comes from relatively faint sources and this
quantity is essentially constant for large enough values of $S_d$.

Clustering accounts for 10--20\% of the total confusion noise,
depending on the assumed model for $\xi_{\rm sph}$, for the
$160\,\mu$m MIPS channel and the SPIRE channels, but its
contribution increases up to 35--40\% of the total confusion noise
for the $175\,\mu$m ISOPHOT channel, and dominates the
fluctuations due to extragalactic sources in the case of
Planck/HFI, except in the longest wavelength channel where Poisson
fluctuations due to radio sources take over. However, higher
resolution surveys can be used to subtract such sources down to
flux densities well below the estimated $S_d$, thus decreasing
their contribution to fluctuations. We note that a similar task is
much more difficult in the case of forming spheroidal galaxies
since the main contribution to their clustering fluctuations comes
from very faint flux densities.

As noted above, the ratio of clustering-to-Poisson fluctuations
increases with decreasing angular resolution. In fact [see
Eq.~(\ref{eq:sigma_poiss})], $\sigma_P \propto \Theta$ while, if
$\xi(r)=(r/r_{0})^{-\gamma}$ with $\gamma=1.8$, $\sigma_C \propto
\Theta^{1.6}$ (De Zotti et al. 1996). It may also be noted that,
as a consequence of the different dependence of $w(\theta)$ on angular
scale (see Fig.~\ref{fig:wth_2mJy_850}), the values of $\sigma_C$
implied by Model 1 (shown in Fig.~\ref{fig:sigma_N}) exceed those
given by Model 2 for higher resolution surveys, while the opposite
happens for lower resolution surveys, such as those of Planck/HFI.

Our estimates -- obtained with the inclusion of the clustering
contribution -- of the confusion noise for the $175\,\mu$m ISOPHOT
channel are in good agreement with the observational determination by
Dole et al. (2001) in the FIRBACK survey fields, once we allow for the
different flux limit. These authors find a confusion noise of
$45\,$mJy when adopting a $3\sigma_{N}$ detection limit of $135\,$mJy,
while we adopt a $5\sigma_{N}$ limit of 327--$350\,$mJy (see
Table~\ref{tab:results}, bearing in mind that $\sigma$ increases with
$S_d$). The possibility of a significant contribution from clustering
in the $175\,\mu$m FIRBACK survey was discussed by Perrotta et
al. (2003) and Dole et al. (2003).

In spite of the different models used and of the different criteria
adopted to define the limiting flux densities, our estimates of
Poisson fluctuations are in reasonable agreement with those by
Rowan-Robinson (2001), Dole et al. (2003) and Xu et al. (2003) for
MIPS/Spitzer and by Lagache et al. (2003) for SPIRE/Herschel and
HFI/Planck. At the wavelengths (160 and 850$\,\mu$m) where the models
are tightly constrained by the data, differences are generally within
20--30 per cent, our predictions being on the high side. At other
wavelengths, our estimates tend to be higher than those of other
authors by up to 50 per cent.

A preliminary attempt to allow for the effect of clustering of
evolving dusty galaxies was carried out by Toffolatti et al.
(1998), who assumed $\xi(r)=[r/r_{0}(z)]^{-1.8}$ and constant
clustering in physical coordinates (stable clustering model:
$r_{0}(z)=r_0(z=0)(1+z)^{1-3/1.8}$, in comoving coordinates). They
further adopted an Einstein-de Sitter universe and set
$r_0(z=0)=10$ Mpc/h. Their model yields a comoving correlation
scale length at $z=3$, $r_0(z=3)\simeq 4$ Mpc/h, i.e. about a
factor of 2 smaller than indicated by current data (see
Sect.~\ref{sec:clustering}). Correspondingly, they have probably
substantially underestimated the clustering fluctuations.

Recently, Takeuchi \& Ishii (2003) have estimated the effect of
clustering with reference to the forthcoming ASTRO-F surveys. They
find that, between 60 and 170$\,\mu$m, clustering increases by
$\sim10$ per cent the detection limit derived by only considering
Poisson fluctuations. Their result is thus compatible with our
estimate for the MIPS/Spitzer 160$\mu$m channel.

Finally, it must be stressed that we have only considered fluctuations
due to extragalactic point sources. Additional contributions of
varying importance (depending on the wavelength and on the position in
the sky of the surveyed area) come from Galactic (synchrotron,
free-free, interstellar dust) and zodiacal emissions. At millimeter
wavelengths the graininess of the sky is actually dominated by
fluctuations of the cosmic microwave background. Further fluctuations
may be produced by Sunyaev-Zel'dovich effects in groups and clusters
of galaxies. Last, but not least, to determine the sensitivity of a
survey we also have to allow for instrumental noise. Therefore, the
values of $S_d$ quoted in Table~\ref{tab:results} must be regarded as
lower limits.

\noindent
\section*{ACKNOWLEDGMENTS}
Work supported in part by ASI and MIUR. We warmly thank G.
Rodighiero for useful discussions and the anonymous referee for
comments that helped improving the paper.

\end{document}